\documentclass[10pt]{article}
\usepackage{graphicx}% Include figure files

\usepackage{epsfig}
\usepackage{latexsym}
\usepackage{amsmath}

\begin{document}

\title{  %Measure of the
Binary Scientific Star Coauthors Core Size   %\\    mav09jpeg
}

\author{Marcel AUSLOOS$^{1,2,}$%$^{a}$
\footnote{{\it previously at}
GRAPES@SUPRATECS, ULG,   Sart-Tilman, B-4000 Li\`ege, Euroland}   \\ %$^a$
$^{1}$r\'es. Beauvallon, rue de la  Belle Jardini\`ere, 483/0021, \\
B-4031 Li\`ege-Nagleur, Wallonia-Brussels Federation \\       $^{2}$ "Affiliated researcher",  eHumanities group\\
Royal Netherlands Academy of Arts and Sciences\\
Joan Muyskenweg 25, 1096 CJ Amsterdam, The Netherlands 
\\ %{\it previously at}\\ 
%GRAPES@SUPRATECS, ULG,   Sart-Tilman,\\ B-4000 Li\`ege, Euroland
\protect\\ e-mail address: marcel.ausloos@ulg.ac.be }

 \date{\today}
\maketitle
 \vskip 0.5 cm

\begin{abstract}
It is examined whether the relationship  $ J \propto A/r^{\alpha}$, and the subsequent coauthor core notion (Ausloos 2013), between the number ($J$) of joint  publications (JP) by  a  "main scientist"  (LI) with her/his coauthors (CAs)   can be extended to a team-like system.   This  is done by considering that each coauthor   can be so strongly tied to the LI  that they  are forming {\it binary scientific star} (BSS) systems with respect to  their other collaborators. Moreover,  publications in peer review journals and in "proceedings", both often thought to be of "different quality", are separetely distinguished.  The role of  a  time interval for measuring  $J$ and $\alpha$  is also examined. New indirect measures are  also introduced.
 
  For making the point, two  LI cases with numerous CAs are studied.    It is    found that only a few BSS need to be usefully examined. The exponent $\alpha$ turns out to be "second scientist"  weakly dependent, but still "size"  and "publication type" dependent, according to the number of CAs or JP. The  CA  core value is  found to be (CA or JP) size and  publication type dependent, but remains in an understandable range. Somewhat unexpectedly, no special qualitative difference on the binary scientific star CA core value is found between  publications in peer review journals and in proceedings.  
  
  In conclusion, some remark is made on partner cooperation in  BSS teams. It is suggested that such measures can serve as criteria for distinguishing the role of scientists in a team. 
 \end{abstract}

 \section{Introduction  }\label{sec:intro}

 In (Ausloos 2013), it was found out that   a  Zipf-like law
 \begin{equation}   \label{eqZipf}  J \propto 1/r ,
  \end{equation} 
  exists,  between 
the number ($J$) of  joint publications of a  "main scientist", called for short "leading investigator" (LI) with her/his coauthor  (CA),  when both  the number  $J$  and   of CAs are   "large"; $r $ (=1,... )  is an integer allowing some hierarchical ranking of  the CAs; $r=1$ being the most prolific CA with the LI, such that $r_M$ is the number of CAs of a LI.
 
  However, it was  observed that  a hyperbolic  (scaling) law   is more appropriate
 \begin{equation}    \label{eq1}
 J =  A/r^{\alpha} ,
  \end{equation}
 with $\alpha\neq1$;  usually  $\alpha\le1$.
 
  This finding  allows one to define  the {\it core of coauthors of a scientist} through  a  threshold (Ausloos 2013),
 called  the  $m_a$-index, which takes  the  largest $r$  possible value
  % \begin{equation} \label{eq2}
($   m_a \; \equiv \; r),$ %    \;  \;  \;  \;  $
%$as   \; long \; as  $
 such that $ r\;$  $  \le \; J.$
 % \end{equation}
   This index measures the core of the most relevant  coauthors in a research team,   centered on  the  LI,. This threshold  definition is analogous  to that defining the $h$-index  (Hirsch 2005, Hirsch 2010). Recall that the latter  is a measure of the {\it core of papers} of an author,  - some appropriate, though debatable,  "best output" measure  (Rousseau 2006, Kelly \& Jennions 2006).   In this $h-$index method,  one implicitly assumes that the number of "important papers" of an author, those which are the most often quoted, allows to measure the impact of a researcher  (Durieux \& Gevenois 2010). However, such a citation effect is often due to the activity of  a research team,   centered on  the  LI   (McDonald 1995, Melin \& Persson 1996, Kwok 2005). In fact, there  has been much work on "improving"  the $h$-index  (Jin 2006, Jin 2007, Jin et  al. 2007, Persson et al. 2004, Bornmann  \& Daniel  2009,  Zhang 2009),  e.g., for taking into account multi authored papers, journal impact factors, etc., thereby leading to many variants (Bornmann et al. 2008, Schreiber 2010, Schreiber et al. 2012).

     In contrast, the $m_a$ index (Ausloos 2013) measures the role of coauthors, rather than citations, and indicates the most important coworkers of a LI.  Technically, one could thus measure  both the relevant size and the so called strength of a research group, centered on some leader, thereby measuring  some impact of research collaboration, e.g.,  on scientific productivity  as studied in (Lee \& Bozeman 2005).   The invisible college  (Kretschmer 1994, Zuccala 2006)   of a LI would become visible, easily quantified, whence pointing out to some  criterion for some selection in the community.

Moreover, it is of common knowledge that a LI often delegates some responsibility to senior researchers in order to pursue some activity in specific fields,  - often, though not always,  remaining in charge of each publication  (Kretschmer 1985). Furthermore, there are other reasons why a LI has CAs.  It can be easily thought and even argued that a LI increases his/her ease in publishing because of the many CAs. Officially,  co-authorship should  imply personal responsibility for the content of  a paper. However,  some   "team reputation" in  having  either active (senior)  investigators   or (and)  to show a large set of  (junior) CAs cannot be neglected. It should be easily admitted that the (junior or senior) CAs are not full of altruism. Publishing with a well known LI brings some attention, and can be likely useful for a  career     (Petersen et al. 2011, Zhou et al. 2012).  Conversely, some sense of "obligation",  need for justifying some collaboration, pressure from another coauthor, or explicit demand, gaining favor or reciprocity, fear of offending someone or searching for some potential grant, are common factors.  In view of these, one has to shock readers by recalling Hollis observation  (Hollis 2001) of a negative relationship between collaboration and output.    Hollis  showed  that for a given individual, more coauthorship is associated with higher quality, greater length, and greater frequency of publications. However, the net relationship between coauthorship and net output attributable to the individual is negative, i.e. after discounting for the number of authors. This raises the question for the need, the role,  the quality, the quantity of  coauthors, and what a core of CAs  truly means.

Yet,   undeserved authorship, assigning authorship to persons because of their authority or prestige, or as courtesy,  seems much allowed: the percentage of undeserved (false) authorship  has been shown to  increase   (Slone 1996,  Vuckovic-Dekic 2000)   along with the increasing number of coauthors.  
  
Moreover, it  is somewhat  commonly accepted that  proceedings papers, e.g. resulting from conference presentations,  are less "valuable", more easily published, and contain more coauthors than peer review journal publications.
   Whence, Miskiewicz  (Miskiewicz 2013) recently discussed whether  publications through peer review journals  in contrast to so called "proceedings"\footnote{Proceedings usually contain papers resulting from presentations at scientific meetings. By including into "proceedings" (with quotation marks) papers published in encyclopedia, in vulgarization journals, or as chapters of books $not$ related to scientific meetings, one is allowing some reasonable statistical analysis by including otherwise outliers into a sound category of papers.} have some impact on the core number and on the  ranking of CAs. He found also some indication concerning time intervals in a  publishing list,  though   such differences might be attributed to the new electronic publishing means.
  
  On the other hand, it frequently occurs  that a theoretician is publishing with experimentalists and conversely, - whatever the scientific field.  In that line of reasoning, Bougrine   (Bougrine 2014)  in a remarkably thorough paper sorted out the subfields of several scientists and discussed the subfield effects on the $m_a$  cores.
 
 Therefore,   a complementary question to the above investigations   is hereby examined, i.e. whether a  "binary scientific star"-like system is also implied  in   Eq.(\ref{eq1}), or whether some deviation occurs.  Thereafter, can one debate within the new above framework on such cooperation  states?  
 
 The  "binary scientific star"   (BSS) is defined as  the couple formed by the LI and one of his  CAs. Their joint publications, with of course other CAs, is here examined in the spirit of studies on the "core of co-authors" (Ausloos 2013).  
 In line with previous publications (Ausloos 2013, Miskiewicz 2013, Bougrine 2014), two active LIs list of publications with  their CAs are specifically examined.  For both LIs, scientific papers in peer review journals are distinguished from, but are counted with the same weight as,  publications in proceedings and similar media, like encyclopedia, summer school lecture books,  other book chapters, etc., - thereafter called "proceedings".  %Whatever the number of authors, their rank in the list of CAs for a given paper,  the paper citation frequency, the impact factor of the journal,  the number of pages or characters, etc.

 Deviations from Eq.(\ref{eq1}) are going to be found to occur at low rank  as in   (Ausloos 2013, Miskiewicz 2013, Bougrine 2014). Both a king effect   (Laherr\`ere \& Sornette 1998)  and a queen effect  (Ausloos 2013)  may exist, as indicated by an upturn at $r\le1$ and a   horizontal curbing for $r\le 2, 3, ...,$ respectively,   in a log-log plot of  the data.  This at least allows to visualize the main coauthors of  a LI and allows one to restrict at once  the scope of the present  investigations of  "BSS" systems to a few CAs only, - fortunately.    
  Various power law exponents $\alpha$ are found for Eq.(\ref{eq1}).  The $m_a$ value of the core of "BSS coauthors" is also  examined and deduced from the numerical analysis. 
  
  Note that to look for CAs of a  LI is nothing else that  to measure the degree $k_i$  (the number of connections of a vertex)  of the LI as a node $i$  in his/her  scientific  collaboration network   (Newman  2001a, 2001b). 
On the other hand, to search for the distribution of CAs for a BSS is equivalent to obtaining the number of triads containing  one specific link in a collaboration network.  It should also appear at once that the {\it number of different coauthors} (NDCA)    is  equal to the highest possible rank  value, $r_M$.  Moreover,  in  the case of a   BSS, and only in such a case, NDCA and $r_M$ are  identical to the number of triads. Watts and Strogatz   (1998)  pointed out that most networks are highly ÒclusteredÓ, in the context of social networks. This means that there  is a heightened density of ÒtrianglesÓ of acquaintances in the network.  A histogram
of the degrees of vertices in a real-world network often indicates a  highly
skewed distribution.  Whence,  the skewness ($skw$) and the kurtosis ($krt$) of the distributions  of CAs of such BSSs will be discussed.    
   
     The methodology is  briefly explained in Sect. \ref{sec:Method}. The data analysis of the  coauthorship features is contained in Sect. \ref{sec:data}.   In Sect. \ref{sec:dataanal}, a few indirect measures are presented and discussed.
In Sect. \ref{sec:dataset}, some discussion on the statistical aspects of these illustrative cases are presented in line with general considerations   for a LI role. Sect. \ref{sec:conclusions} serves as a conclusion with some suggestion for future work, like removing constraints in the present approach, and  imagining applications.

     \begin{table}
     \caption{  NP: total number of publications; NJP: number of joint publications with CAs; NJPj  in  journals (j); NJPp in so called  "proceedings' (p), i.e. including   joint publications in book chapters and in encyclopedia chapters;   updated  in Dec.  2012;   the number of different co-authors is given as NDCA, also for that corresponding to peer review journals, NDCAj, and  proceedings NDCAp % sA : single author;
 }\label{datasummary} \begin{center}
      \begin{tabular}{|c| |c|c|c|c|c|c|c|c|    }
  \hline
&  \multicolumn{7}{|c|}{publication data summary}   \\
\hline     & NP&NJP&NJPj&  NJPp&NDCA&NDCAj&NDCAp%&NsAP&NsAPp%&NJPbc&NsAPbc&NJPe&NsAPe&  
  \\
\hline     HES& 1150&1092&791&301&592&565&242%&(15)&(43)%&9&6&2&3&?  
  \\
% \hline  DS&&599&359&29&148&31&15&5&1&1&?   \\ 
 \hline  MA & 599&523&359&164&319&273&172%&(29)&(37)%&15&5&1&1&? 
   \\ 
\hline   
% \hline  \bf{JM} &&NP&NJPj&NsAPj&NJPp&NsAPp&NJPbc&NsAPbc&NJPe&NsAPe&?    \\ 
 \hline  HES\&SH& 300&299&195&104&196&169&114%&(29)&(37)%&15&5&1&1&? 
   \\ 
    \hline  HES\&SB & 262&262&164&98&216&175&124%&(29)&(37)%&15&5&1&1&? 
   \\ 
    \hline  MA\&RC & 155&155&97&58&147&114&71%&(29)&(37)%&15&5&1&1&? 
   \\ 
    \hline  MA\&NV & 89&88&60&28&40&28& 24%&(29)&(37)%&15&5&1&1&? 
   \\    \hline  \end{tabular}   \end{center}
\end{table}

\section{Methodology} \label{sec:Method}

For the following study and discussion  the   considered  LIs  are  among those investigated in several previous publications  (Ausloos 2013, Miskiewicz 2013, Bougrine 2014). They are called HES (H.E. Stanley) and MA (M. Ausloos) for short. 

HES publication list amounts to more than 1100 "papers". Its  joint publications with  (many, $\sim $ 600 ) CAs is taken from  $polymer.bu.edu/hes/vitahes-messina.pdf $.   The list has been broken into four classes: two  are distinguishing peer review papers in scientific journals (j) from a list including (15) book chapters,  5 encyclopedia articles and papers resulting form scientific contributions at conferences, thus, see previous footnote, so called "proceedings" (p).  
MA has   a list of about 600 publications, with  $\sim $ 300 CAs, mixing papers in peer review journals, proceedings,   encyclopedia and book chapters.  

This data, updated  in Dec.  2012,  is concatenated  in Table 1 (top), i.e.  NP is  the  total number   of publications; NJP: the number of joint publications (JP), i.e., $ J$ in Eqs.(\ref{eqZipf})-(\ref{eq1}); broken into NJPj, i.e.,  those  in  journals (j); NJPp, those  in so called  "proceedings" (p),  thus somewhat  adapting   (or "generalizing") the notion of proceedings, i.e. including   joint publications in book chapters and in encyclopedia chapters. In the Table, one also finds   the number of different co-authors, i.e. NDCA, whence $r_M$. The  notations  are rather obvious:   the publications in   peer review journals or "proceedings" are called NDCAj or NDCAp, respectively.  %sA : single author; 

 Sometimes,  some ambiguity occurs on whether a publication pertains to a scientific report presented at a conference or is a truly more elaborated paper with original findings not yet published in a peer-review journal. Indeed, several proceedings appear in scientific journal special issues.    Sometimes several contributions are not truly scientific papers reporting original work,  but contributions in view of vulgarization.  These often involve CAs. In view of this, I took them into account as well. However, identical scientific publications  have $not$ been counted twice: for example, HES has many papers reproduced,  e.g. in a compendium or in another journal.   %Note that in Table 1, the single author  (sA) papers are distinguished from the joint publications (JP), either in peer review journals  or in the general proceedings category.  

  Moreover, HES distinguishes between (j and p) publications before and including, 1999 from those after,  and including,   2000.    It was easy to break the MA publication list into such two  time intervals,  identical to those of HES lists,  as well, also distinguishing between j and p, in these time intervals.  The notations, j1, j2,  p1 and p2 are subsequently used for distinguishing time intervals. They can be combined into  j1p1, j2p2, and into  j1j2 $\equiv$ j,  p1p2 $\equiv$ p,   and j1j2p1p2 $\equiv$  jp.   
Thus 18 data sets are available for study. Observe that the "first" time interval spans about 30 years, the "second" about 10 years.
   
  Later on, CAs have been counted manually and visually  in each list, using   a web engine
   $http://rainbow.arch.scriptmania.com/tools/word_{-}counter.html$.    Great care, -  a tedious work,  was taken about CAs misspelling, and  sometimes changing of name or initials with time, - in order to fully count the contribution of each CA, and not distribute it on several persons, - who are in fact only one. 

\begin{table}  \begin{center} 
\caption{Number of joint publications (NJP) of HES and MA, further broken into  journals (j) or so called "proceedings" (p),  with their 5 main CAs, thus making a few BSS systems %; NDCA : number of different CAs  in each case 
 }\label{Tablestat}
 \begin{tabular}{|c|    |c|c|c|c|c|     |c|c|c|c|c|  }
   \hline
&\multicolumn{5}{|c|}{\bf{HES \&}}     &
 \multicolumn{5}{|c|}{\bf{MA \&} }  \\

    &  SH& SB& LA &FS&CKP& RC & NV&PV&AR&HB    \\ \hline 
  NJP &299&262&87&73&73& 155 &88&63&	62&	61    \\
NJPj&195&164&40&48&37&97&60&42&39&32\\
NJPp& 104&98&47&25&36&58&28&21 &23&29\\
  %NDCA???&196v&216v&???&73&73& 147v &40v&63&	62&	61    \\
%NDCAj???&169v&175v&40&48&37&114v&28v&42&39&32\\
%NDCAp???&114v&124v&47&25&36&71v&24v&21 &23&29\\
 \hline
\end{tabular}

 \end{center}
 \end{table}

   \section{Display  of the Data Set Numerical Content}\label{sec:data}  
   
   Having established the  rules for gathering the various data to be analyzed, and obtained the latter, 
 it seems relevant to outline scientific questions of interest emphasizing the numerical aspects used for answers. Their discussion will arise in Sect.5.
  
 \begin{itemize}
 \item Crude statistical aspects.
 
 First  for completeness and for  further reference, let   the  statistical characteristics of the distributions of  joint publications with all CAs,  for HES and MA, as LI, both in journals (j) and in so called ÓproceedingsÓ (p),  and their "sum" be displayed. This is reported in Table 2, together with the deduced  $\alpha$ 	and $m_a$ values, and  their NDCA ($\equiv r_M$). 

\item Number of "interesting" BSS.

In order to limit the number of interesting BSSs to be investigated, it seems reasonable only to consider the CAs mainly with low rank, within the king and queen effect regimes, thus much below the $m_a$  measure of the LI core.   Such CAs are listed in Table 3, with a few characteristics publication  data, i.e.  for $r= $ 1 to 5,  for HES, Havlin (SH), Buldyrev (SB),   Amaral (LA),  Sciortino (FS), Peng (CKP); they are followed by  Ivanov (PCI), Goldberger A. (AG), Gopikrishnan (BG),  and Plerou (VP),  for which data is not listed here,  for  space savings, but they have been studied. Similarly, for MA, Cloots (RC), Vandewalle (NV), Vanderbemden (PV), Rulmont (AR),  and Bougrine (HB)  get  a $r= $ 1 to 5,  respectively, see Table 3. %

\item Outstanding BSS?

In both LI cases, all these  low ranking CAs have   a large number of joint publications (NJP), sometimes sharing some,  of course.  For both LIs, however,  as it readily appears, from Table 3, two CAs much stand up over the others, i.e. their NJP $\ge 260$ or $\ge 85$, respectively,  much above those with $r  \ge$ 3 . Thus, only such two prominent CAs are investigated in the present illustration of  BSS  systems, i.e. SH and SB on one hand, for HES, and RC and NV, on the other hand for MA. 

\item Detailed statistics and parameters to be found.

Thereafter, the selection of  such 4 BSS cases, i.e. s  HES\&SH, HES\&SB, MA\&RC  and   MA\&NV has been made for illustration and discussion.  For completeness, the detailed data on their NP, NJP and NDCA is given in Table 1 (bottom), for these 4 BSSs. The characteristics of the distributions of each 4 BSSs, only for their whole set (jp) of joint publications with their respective CAs (and LI),  are given in Table 4. The best power law fits, through Eq.(2),  lead to $\alpha$ with some $R^2$;  each $m_a$ value is reported. The distribution statistical characteristics are given. 

\item Differences and similarities to be examined. Time dependence effects.

Tables 5-8 are summarizing data previous to questions on similarity in behavior of the BSS.,  They visually  point to several differences to be examined next in more detail through the different subsets,  see Sect. \ref{sec:dataset}.  In particular, a time dependent effect on $\alpha$ and $m_a$ should be examined, in both  types of publications here above defined.

   \end{itemize}

%&\multicolumn{3}{|c|}{DS}

     \begin{table}
     \caption{  Characteristics of  the distributions of HES and MA, as LI,  joint publications  with all CAs,  in  journals (j) and  in so called  "proceedings" (p), i.e. including   joint publications in book chapters and in encyclopedia chapters, and the whole set (jp);   updated  in Dec.  2012. Best power law fits, Eq.(\ref{eq1}),  $m_a$ value, and statistical characteristics are given, illustrating the similarity in behavior %of the data and subsequent analysis
  }\label{HESMAbasics1} \begin{center}
      \begin{tabular}{|c| |c|c|c|c|c|c|c| c| c| c|     }
  \hline
&  \multicolumn{10}{|c|}{LI distribution characteristics of CAs}   \\
\hline     && $A$&$\alpha$&$R^2$& $m_a$&$\sum$&$\mu$&$skw$&$krt$&NDCA%&NJPbc&NsAPbc&NJPe&NsAPe&  
  \\
\hline
\hline    HES&j&473.41&0.999&0.914&20&2639&4.671&11.42&161.15&565%&9&6&2&3&?  
\\ 
\hline    HES&p&275.42 &1.045&0.875&15&1243&5.136&6.281&47.86&242%&9&6&2&3&?  
  \\
  \hline    HES&jp&1200.57&1.135&0.868&26&3889& 6.569&11.097&153.42&592%&9&6&2&3&?  
  \\
 \hline
 \hline  MA &j&205.82&1.029&0.91&15&1055&3.865&6.508&55.1&273%&15&5&1&1&? 
   \\ 
    \hline  MA &p&62.65&0.859&0.981&10&523 &3.041&6.020&44.40&172%&15&5&1&1&? 
   \\ 
    \hline  MA &jp&380.28&1.102&0.903&19&1554&4.872&7.371&70.40&319%&15&5&1&1&? 
   \\ 
\hline 
 \end{tabular}   \end{center}
\end{table}

  \section{A few Indirect Measure Definitions }  \label{sec:dataanal}

As in the $h-$ index discussions, criticizing the reduction of a scientific impact, through publications citations, to a mere scalar number,  it  can be argued that the $m_a$  index has in itself the same type of defect, - reducing a team of coauthors impact to a scalar number. Therefore, along the lines of development  and introduction of other indices following the $h$-index appearance, and in order of preventing such criticisms, one can also suggest other practical measures, in the CA-core index notion, considering parts or the whole  $J(r)$ histogram.  This would allow some some "vectorial comparison" of scientists or teams when necessary.

%For example, several  measure definitions can  take  into account    either  (i) the whole surface  or (ii) part of the  $J$ $vs.$ $r$ histogram. 
Indeed, the   whole cumulated NJP is 
 \begin{equation}\label{sum}
\sum  \equiv \sum_{r=1}^{r_M} J_{r}  ,
\end{equation}
 in the case that the CA with rank $r$ has published $J_r$ publications with the LI, or with the BSS, - depending on what  case is  at hand. 
  Part of  the histogram can also be examined, i.e. that cumulated NJP {\it  limited to the core,} as 
 \begin{equation}\label{suma}
A_{a}   \equiv \sum_{r=1}^{m_a} J_{r}  .
\end{equation}
The notations are reminiscent of  the $A-$index (Jin 2006),  in the Hirsch scientific output measurement method.  

Similarly to the $h$-index literature, one can define   relative indices which measure  the  whole surface below the empirical data of the number of joint publications, i.e., 
  \begin{equation}\label{a_aa}
a_M  =\frac{ 1}{m_a} \sum_{r=1}^{r_M} J_{r}\; \equiv \; \frac{\sum}{m_a},
  \end{equation}
  and  {\it  till the CA of rank } $m_a$,  i.e., the l.h.s. part of the histogram, 
\begin{equation}\label{a_a}
 a_a =\frac{ 1}{m_a} \sum_{r=1}^{m_a} J_{r}\; \equiv \; \frac{A_a}{m_a}.
  \end{equation}

Of course  $A_a/\sum$, see Tables 6-8,   $\equiv a_a/a_M$.  Obviously, $A_a/\sum$ gives the relative weight of the core CAs in the cumulated NJP of a LI  or a BSS.  Another indirect  index  
  \begin{equation}\label{mu}
 \frac{ 1}{r_M} \sum_{r=1}^{r_M} J_{r}  \; \equiv \; \frac{\sum}{r_M}.
  \end{equation}
is nothing else that the mean $\mu$ of the distribution\footnote{Recall that NDCA is identical to the maximum possible rank $r_M$, the value of $M$ depending on the case at hand}, i.e. the average number of JP  per CA.
  It ranges between 2 and 6 (see Table 5). Resulting data  and statistical  characteristics are summarized in Tables 3-8.

     \begin{table}
     \caption{  Characteristics of  the distributions of  each whole set  (jp) of  joint publications  by 4  BSSs, HES\&SH, HES\&SB, on one hand, and MA\&RC, MA\&NV, on the other hand,   with all their CAs;   updated  in Dec.  2012.  The best power law fits, Eq.(\ref{eq1}),  leads to  $\alpha$ values with some $R^2$; the deduced $m_a$ value from a $J$ $vs.$ $r$ plot is reported;  statistical characteristics are given, illustrating the similarity in behavior %of the data and subsequent analysis 
 }\label{HESMAbasics2} \begin{center}
      \begin{tabular}{|c| |c|c|c|c|c|c|c| c| c| c|     }
  \hline
&  \multicolumn{10}{|c|}{LI distribution characteristics of CAs}   \\
\hline     && $A$&$\alpha$&$R^2$& $m_a$&$\sum$&$\mu$&$skw$&$krt$& NDCA%&NJPbc&NsAPbc&NJPe&NsAPe&  
  \\
\hline
\hline    HES\&SH&jp &272.12&1.074&0.946&16&1098&5.602&7.309&68.91&196%&9&6&2&3&?  
\\ 
 \hline    HES\&SB&jp &260.20&1.064&0.925&15&1088&5.014&8.388&92.48&216%&9&6&2&3&?  
  \\
 \hline
 \hline  MA\&RC &jp &99.90&0.985&0.883&11&560&3.810&4.417&22.09&147%&15&5&1&1&? 
   \\ 
 \hline  MA\&NV &jp&16.35&0.835&0.860&5&104&2.6&5.329&28.89&40%&15&5&1&1&? 
   \\ 
\hline 
 \end{tabular}   \end{center}
\end{table}

  In practical terms,   these indirect measures are  attempts to improve the sensitivity of the $m_a-$index to take into account the number of  relevant  co-authors,  whatever the number of joint publications among  the most frequent coauthors, whence to introduce a contrast between the most frequent CAs and the less frequent ones.  Indeed in the fits, through Eq.(\ref{eq1}), the influence of accidental or rare CAs,  i.e. with large $r$ values,  can be rather huge for  estimating the amplitude $A$ and the exponent $\alpha$ in Eq. (2).

  \section{  Discussion  }\label{sec:dataset}

     Several  possible ways exist in order to display   and to analyze  the whole data. Some difficulty arises because, as often, questions  are intertwined; so are the answers.
     
     \begin{itemize}
     \item  One can observe whether Eq.(\ref{eq1}) holds for the 4  BSS cases when p1, p2, j1 and j2 are taken independently of each other. \item One can  examine pairs like p1p2 and j1j2, i.e. comparing the  evolution of scientific contributions $types$, in the j or p categories. \item One can   examine the pairs j1p1 and j2p2, thereby examining some $time$ evolution, whatever the publication type. \item One should surely examine the case jp which encompasses all others, on both grounds, types and time. 
     \end{itemize}
     
      It is obvious that a  display of each item would lead to an enormous amount of figures. Instead, the whole data is summarized in Tables  4-5   and only a few cases serve as an illustration through Figs. 1-7.   They have been selected for their interest in showing excellent or bad fits, through the  resulting $R^2$ value, and to illustrate at least once each of the 4 BSSs. Many other figures are available from the author upon request.

%The notations jp is a short hand notation for j1j2p1p2 (or j1p1j2p2).

  From a more general point of view, one  may begin  discussing the 2 parameters appearing through the power law, Eq.(\ref{eq1}), i.e. $A$ and $\alpha$. It is usual to consider whether the fits are valid if the correlation coefficient $R^2$ has a large value (e.g., $\ge0.9$).  It is hereby confirmed that a sufficient number of CAs or JP must be examined for a given LI,  if some meaningful aspect is derived from  the fits. Thus, several MA\&NV cases are not further examined, and their data is   not  willingly reported in Table 5. 

 %\subsection{Time interval effects}\label{timeinterv}   \subsection{Distribution characteristics: skewness and kurtosis}\label{skkrt} \subsection{Cores and Out-of-cores CAs}\label{COoC}

Therefore, in the following,   \begin{itemize} \item results pertaining to the time interval effect are first  commented upon in Sect. \ref{timeinterv}, 
 
\item followed by
  the statistical characteristics of the CAs distributions in Sect. \ref{skkrt}; \item  the resulting core values, and finally  \item the related "indirect"  measures are discussed in Sect.\ref{COoC}
\end{itemize}

\subsection{Time interval effects}\label{timeinterv} %  \subsection{Distribution characteristics: skewness and kurtosis}\label{skkrt} \subsection{Cores and Out-of-cores CAs}\label{COoC}

  As a  first pertinent question, let it be observed whether some stability exists as a function of time in the behaviour of BSS teams. Recall that  the "first" time interval spans about 30 years, the "second" about 10 years.  A summary of the  analyses and results comparison can be found in the Tables. Some thorough examination leads to   qualitative and quantitative  observations. 
Whatever the time interval, one has  \begin{itemize} 
      \item   a similar  hierarchy holds, on all measures,  for HES\&SH with respect to   HES\&SB as  for MA\&RC with respect to  MA\&NV
       \item the amplitude $A$ well reflects the importance of the relative number of joint publications in all cases, whatever  any occurrence of the king and queen effects.
        \end{itemize}
 
  Differences may occur depending on the time interval : \begin{itemize} 
      \item the exponent $\alpha$ is usually close to 1, but is slightly larger for HES than MA; however, it markedly depends on the NJP, since in a few cases it can fall to $\sim$ 0.7,  see the p2  HES\&SH case (Table 5) , or much below, in cases not shown, because the fits are not sufficiently $R^2$ meaningful;  yet see  the MA\&NV case in Fig. 5  %(or Fig.6) 
  as such an illustration
   \item the exponent $\alpha$  for either p1 and p2, and for j1 and j2, for HES\&SH and HES\&SB  is always larger,  though  not much,  than for MA\&RC
   \item  the exponent $\alpha$   for the time interval 1, i.e. the set p1j1, is larger than the $\alpha$ value for the set p2j2, 
   \item   the exponent $\alpha$   for  p1 is smaller than for the set p1j1,  as well as smaller than for j1
   \item  the exponent $\alpha$   for  p1 is  larger than the $\alpha$ value for the set  j2
   \item    the exponent $\alpha$ for  p1, for j1, for p1p2, an for j1j2  has a systematic hierarchy, for the BSS,    as if  there was a ranking HES\&SH, MA\&RC,  HES\&SB
 %  \item  except j2 and p2j2, {\it for which the ranking is reversed.??}
 %  \item  ...
  \end{itemize}
  
  Therefore, the main conclusion from observing  such  time intervals    and publication subsets leads to confirm the importance of the relative number of joint publications in all  BSS cases,  within a finite time interval, for obtaining $R^2$ meaningful values. Also the LI scientific output differences  play a relatively similar role on the  parameter values. Hierarchies are confirmed in each  time interval for all BSS  cases. Apparently these time interval studies  do not indicate a JP behavior departure for the LIs or BSSs as a function of time.
  
  \subsection{Distribution characteristics: skewness and kurtosis}\label{skkrt} %\subsection{Cores and Out-of-cores CAs}\label{COoC}
  
 Some statistical analysis is meaningful if the distributions are not too anomalous. Therefore it is of interest to discuss higher order moments than the second.  Several meaningful observations can be related to the characteristics of the distribution of joint publications with CAs, as given in the right hand side of Tables 4-5.  One should observe anomalous and regular trends. For example,
  
    \begin{itemize}
  \item    the $\mu$  values   have a short range,  from $\sim 2$ to $\sim 6$,  for each LI and BSS, whatever the subset
   \item  the  $skw$  values fall into a  small  range, about equivalent for the HES or MA cases:  roughly ]2; 5[, however, with an anomalous value  $\sim 7.2$ for j1j2 of HES\&SH
 \item whence implying a large $skw$ value for the jp, i.e. $\sim 8$
     \item the $skw$ for a BSS seems to be $\sim$ 80\% of the corresponding LI value
        \item the $krt$ increases much with the number of JP, culminating in  a value  $\sim83$  for j1j2 of  HES\&SB 
    \item  the $krt$ ranges also much differ for HES and MA,  going from  $\sim 6$  till  $\sim 36$  for HES, but only from  $\sim 3$ till  $\sim 24$   for MA; - see Table 5
       \item whence implying a large $krt$ value for the jp, i.e. $\sim 90$  for  HES\&SB, - see Table 4
       \item and     $\sim 30$  for MA\&NV, - see Table 4.
            \item the $krt$ for a BSS seems to be $\sim$ 1/3  of the corresponding LI value.
  \end{itemize}
   
   Observe that the  (total, i.e. jp) distribution  kurtosis   for  the  two main BSSs with HES relative  to that of the MA-BSS is very large. This is essentially due to the  relatively larger NDCA in peer review journals, for the former LI, -  see last  line  of Table 5.
   
Motivated by observations and arguments by Price (1956) and  others (Fern\' andez-Cano et al. 2004),  it has been searched whether the evolution of $skw(r_M)$ and  $krt(r_M)$  is close to  an exponential grow (or to a power law in case of scaling) for either HES and MA, but also for HES\&SH, HES\&SB, MA\&RC, and MA\&NV.  In all cases, one has $R^2\sim 0.8$. However in view of the limited amount of data,  and performing  a Jake-Berra test, neither analytic form is convincing.  It is observed that the mean of the distribution has an erratic behavior as a function of  NJP  or NDCA (i.e., $r_M$),  then leading to a very low $R^2\sim 0.5$. Yet,  $\sum$   has a  fine behavior as a function of  NJP  or NDCA (i.e., $r_M$),  then leading to a  $R^2\sim 0.9$.  The worse and best cases are shown in Fig. \ref{fig:Plot37musum4fitslilipdf}.

\begin{table}\caption{ Time interval dependence of the CA core value and other characteristics of the number of joint publication distributions in peer review journals for the 4 BSSs so examined in the text; $a_a= A_a/m_a$; $a_M = $$\sum$/$m_a$; recall that the time interval 1 and 2 are about 30 and 10 years respectively
 }\label{Tablestat6}  \begin{center} 
 \begin{tabular}{|c|    |c|c|c|c|c|c|c|c|c|c|c|c|  }
  \hline  
  &\multicolumn{2 }{|c|}{(HES \& }  & \multicolumn{2 }{|c|}{(MA \& }& \multicolumn{2 }{|c|}{(HES \& }  & \multicolumn{2 }{|c|}{(MA \& }   &  \multicolumn{2 }{|c|}{(HES \& }  & \multicolumn{2 }{|c|}{(MA \& }   \\
%\hline  
  & SH)& SB)& RC)& NV)  
 & SH)& SB)& RC)& NV)  
  & SH)& SB)& RC)& NV)    \\
\hline
%  \hline
&\multicolumn{4}{|c|}{\bf{j1}}    &
 \multicolumn{4}{|c|}{\bf{j2} }& \multicolumn{4}{|c|}{\bf{j1j2} }  \\
%\hline   
\hline   
$m_a$ &9&7&7&3&7&8&5&-&11&10&9&- \\
$r_M$ &98&74&60&23&95& 127 &64&-&169&175 &114& -  \\
$A_a$& 156&109&93&25&98&110&60&-&236&102&151&- \\
$\sum$& 370&243&191&51&284&378&159&-&654&621&350&- \\
$A_a/\sum$&0.422&0.449&0.487&0.457&0.345&0.291&0.377&- &0.361&0.164&0.431&- \\ 
$a_a$& 17.33&15.57&13.29&8.33&14&13.75&12&- &21.45&10.2&16.78&\\
$a_M$& 41.1&34.7&27.3&17&40.6&47.25&31.8&-&59.5&62.1&38.9&- \\  
 \hline
\end{tabular}
 \end{center}
 \end{table}  

 \begin{table} \begin{center}
\caption{  Time interval dependence of the CA core value and other characteristics of the number of joint publication distributions in proceedings for the 4 BSSs so examined in the text; $a_a= A_a/m_a$; $a_M = $$\sum$/$m_a$; recall that the time interval 1 and 2 are about 30 and 10 years respectively
 }\label{Tablestat7}  
 \begin{tabular}{|c||c|c|c|c|c|c|c|c|c|c|c|c|  }
  \hline  
  &\multicolumn{2 }{|c|}{(HES \& }  & \multicolumn{2 }{|c|}{(MA \& }& \multicolumn{2 }{|c|}{(HES \& }  & \multicolumn{2 }{|c|}{(MA \& }   &  \multicolumn{2 }{|c|}{(HES \& }  & \multicolumn{2 }{|c|}{(MA \& }   \\
%\hline  
  & SH)& SB)& RC)& NV)  
 & SH)& SB)& RC)& NV)  
  & SH)& SB)& RC)& NV)    \\
\hline
%   \hline
&\multicolumn{4}{|c|}{\bf{p1}}    &
 \multicolumn{4}{|c|}{\bf{p2} }& \multicolumn{4}{|c|}{\bf{p1p2} }  \\
\hline   
%\hline   
$m_a$ &9&6&6&3&4&9&3&-&10&10&7&3 \\
$r_M$ & 74&66&37&16&60&76 &40&-&114&124 &71& 24 \\
$A_a$& 165&114&69&18&35&105&27&-&204&191&97&19 \\
$\sum$& 321&231&130&35&123&236&80&-&444&467&210&44 \\
$A_a/\sum$&0.514&0.494&0.531&0.514&0.285&0.445&0.340&- &0.460&0.409&0.462&0.432 \\ 
$a_a $& 18.33&19&11.5&6&8.75&11.67&9 &-&20.4&19.1&13.86&6.33\\
$a_M$& 35.7&38.5&21.7&11.7&30.75&26.2&26.7&- &44.4&46.7&30&14.7 \\  
 \hline
\end{tabular} 
 \end{center}  \end{table}

  \begin{table}  \begin{center} 
  \caption{ CA core value and other characteristics of the number of joint publication distributions in peer review journals or proceedings, and for the whole set for the (2) LIs and (4) BSSs, so examined in the text;  $a_a= A_a/m_a$; $a_M =  \sum/m_a$%; see also Table 4
 }\label{Tablestat8}
 \begin{tabular}{|c|    |c|c|c|c|c|c||c|c|c|c|c|c|  }
  \hline  
  &\multicolumn{3 }{|c|}{ HES  }  & \multicolumn{3 }{|c|}{MA }Ž& \multicolumn{2 }{|c|}{(HES \& }  & \multicolumn{2 }{|c|}{(MA \& }     \\
  &\multicolumn{3 }{|c|}{   }  & \multicolumn{3 }{|c|}{  }& SH)&SB)& RC)&NV)      \\
\hline
%   \hline
  &j&p&jp  &j&p&jp &jp&jp&jp &jp     \\
\hline   
%\hline   
$m_a$ &20&15&26&15&10&19&16&15&11&5  \\
$r_M$ & 565&242&592&273&172&319 &196&216&147&40  \\
$A_a$& 895&549&1625&482&221&810&524&469&280&52 \\
$\sum$& 2639&1243&3389&1055&523&1554&1098&1088&560&104  \\
$A_a/\sum$&0.339&0.442&0.479&0.457&0.423&0.521&0.477&0.431 &0.5&0.5\\
$a_a $& 44.75&36.6&62.5&32.13&22.1&42.63&32.75&31.27 &25.45&10.4\\
$a_M$& 132.0&82.9&130.3&70.3&52.3&81.8&68.625&72.53 &50.91&20.8  \\   \hline
\end{tabular}
 \end{center}
 \end{table}

     \begin{table}\label{pjHESMA}
       \begin{center} \caption{ (lhs) Data summary about the number of joint publication distributions of  a BSS made of a  (LI  \& one main CA), in journals (j) or "proceedings" (p), in time interval 1 or 2 (see text); $A$ and $\alpha$ are the fit parameters of Eq.(\ref{eq1}), and $R^2$ the corresponding correlation coefficient. (rhs) Distribution characteristics of  such  joint publications:   
$\sum$ is the number of joint publications, - thus having at least 3 CA;  $\mu$, $skw$, and $krt$ are respectively the mean, skewness and kurtosis of the specific distribution; data updated  from CV of HES and MA  in Dec.  2012
}   \end{center}
      \begin{tabular}{|c|c|c|c|c|c|c|c|c|c|c|c|c|}
  \hline
%\multicolumn{2 }  {|c|}{  } &  \multicolumn{11}{|c|}{ distribution characteristics of publications of  a BSS made of a  (LI  \& one main CA) } &  \hline  
\multicolumn{2 }{|c|}{   }   &\multicolumn{2 }{|c|}{(HES \& }  & \multicolumn{2 }{|c|}{(MA \& }&&  \multicolumn{2 }{|c|}{   } &\multicolumn{2 }{|c|}{(HES \& }  & \multicolumn{2 }{|c|}{(MA \& }    \\
\hline  
 \multicolumn{2 }{|c|}{  }   & SH)& SB)& RC)& NV)&&
\multicolumn{2 }{|c|}{  } & SH)& SB)& RC)& NV)     \\
\hline
\hline   j1&$A$& 74.28&44.77&38.64&8.85&& j1&$\sum$&370&243&191&51    \\
\hline   j1&$\alpha$&0.97&0.94&0.95&0.79&&j1&$\mu$&3.776&3.284&3.183&2.217    \\
\hline   j1&$R^2$&0.932&0.972&0.849&0.854&&j1&$skw$&4.172&5.155&2.685&4.092    \\
\hline   j1&$m_a$&9&7&7&3&&j1&$krt$&22.52&31.89&6.89&15.76    \\
 \hline
 \hline   j2&$A$&40.4&53.5&22.86&-&& j2&$\sum$&284&378&159&-    \\
\hline   j2&$\alpha$& 0.84&0.86&0.815&-&&j2&$\mu$&2.990&2.976&2.484&-   \\
\hline   j2&$R^2$&0.98&0.91&0.95&-&&j2&$skw$&4.806&4.230&3.286&-    \\
\hline   j2&$m_a$&7&8&5&-&&j2&$krt$&29.53&26.49&11.56&-  \\
\hline
\hline   p1&$A$&80.59&44.95&34.46&7.82&& p1&$\sum$&321&231&130&35    \\
\hline   p1&$\alpha$& 1.05&0.963&1.03&0.85&&p1&$\mu$&4.338&3.50&3.514&2.188    \\
\hline   p1&$R^2$&0.87&0.97&0.81&0.895&&p1&$skw$&3.445&4.142&1.966&3.034   \\
\hline   p1&$m_a$&9&6&6&3&&p1&$krt$&12.74&19.00&3.07&8.36    \\
\hline
\hline   p2&$A$& 12.90&40.98&9.875&-&& p2&$\sum$&123&236&80&-    \\
\hline   p2&$\alpha$&0.676&0.891&0.676&-&&p2&$\mu$&2.05&3.105&2.0&-   \\
\hline   p2&$R^2$&0.887&0.821&0.942&-&&p2&$skw$&5.494&2.423&3.596&-    \\
\hline   p2&$m_a$& 4&9&3 &-&&p2&$krt$&34.42&6.29&13.57&-    \\
\hline
\hline   j1p1&$A$&224.65&109.49&76.36&14.23&& j1p1&$\sum$&691&474&322&86    \\
\hline   j1p1&$\alpha$& 1.165&1.059&1.072&0.858&&j1p1&$\mu$&5.906&4.693&4.076&2.606   \\
\hline   j1p1&$R^2$&0.865&0.976&0.786&0.865&&j1p1&$skw$&4.386&5.512&3.120&4.913   \\
\hline   j1p1&$m_a$& 13&9&10&4 &&j1p1&$krt$&23.37&36.05&9.61&23.87    \\
\hline
\hline   j2p2&$A$&65.29&127.1&35.78&-&& j2p2&$\sum$&407&614&239&-    \\
\hline   j2p2&$\alpha$&0.898&0.996&0.87&-&&j2p2&$\mu$&3.67&4.09&2.88&-    \\
\hline   j2p2&$R^2$&0.97&0.856&0.956&-&&j2p2&$skw$&5.884&4.296&4.273&-    \\
\hline   j2p2&$m_a$&9&11&6&-&&j2p2&$krt$&43.81&26.07&20.70&-    \\
\hline
\hline   p1p2&$A$&84.6&88.24&34.92&-&& p1p2&$\sum$&444&467&210&-    \\
\hline   p1p2&$\alpha$&0.974&0.969&0.887&-&&p1p2&$\mu$&3.895&3.766&2.958&-    \\
\hline   p1p2&$R^2$&0.965&0.944&0.879&-&&p1p2&$skw$&5.330&5.710&3.080&-    \\
\hline   p1p2&$m_a$& 10 &10&7&-&&p1p2&$krt$&33.95&42.99&9.30&-   \\
\hline
\hline   j1j2&$A$& 108.4&97.41&53.10&-&& j1j2&$\sum$&654&621&350&-    \\
\hline   j1j2&$\alpha$&0.934&0.922&0.893& -&&j1j2&$\mu$&3.870&3.549&3.070&-    \\
\hline   j1j2&$R^2$&0.95&0.92&0.92&-&&j1j2&$skw$&7.166& 8.010&3.070&-    \\
\hline   j1j2&$m_a$& 11&10&9&-&&j1j2&$krt$&67.48&82.71&16.98&-   \\
\hline \end{tabular} 
\end{table}

   \subsection{Cores and Out-of-cores CAs}\label{COoC}
   
   The main concern of the paper stems in the meaningful existence or not of CA core values through the direct measure $m_a$ and the indirect ones, introduced in Sect. \ref{sec:dataanal}.  Let us  recall  that  these allow to distinguish the CA core and   the whole CA  set contributions.  
   Thus, a discussion of  the $m_a$ for LIs and BSSs, as well as the  $A_a$, $a_a$, $a_M$  values are now in order.   
   
   A few points are readily obvious from the Tables. Therefore,  the few points   outlined here below are those which seem to imply practical considerations or are somewhat unexpected. 
   
   Recall first that $r_M$ (j)  $\in$  (170, 115) $\ge$   $r_M$ (p) $\in$  (120, 50), i.e. there are more CAs in j than in p, - quite contrary to any first expectation.

    Next,  from Table 5, let us first distinguish a few items, i.e. for each BSS, investigating the type of publications:
    
       \begin{itemize}
     \item $m_a$ evolves expectedly when increasing the size of the investigated set, i.e.  $m_a^{(j1j2)}$ $\ge$  $m_a^{(j1)}$ $\ge$ $m_a^{(j2)}$,
     \item and  $m_a^{(p1p2)}$ $\ge$  $m_a^{(p1)}$ $\ge$ $m_a^{(p2)}$,
  \item   interestingly observe a  ratio reversal  for $m_a^{(j2p2)}$/  $m_a^{(j1p1)}$ in the case of HES\&SB, among the other BSSs,  likely due to a different time for SB  in joining the HES group, - as could be  somewhat deduced from Tables 6-7.
    \end{itemize}

Note that the   $m_{a}^{(i)}$ values in  the cases of BSSs are falling  below the   overall $m_{a} $  defining the coauthor  LI-core.  One can usefully compare the values in Table 3 and 4, and observe a factor   roughly  4/3 ($\sim $ $26/15$  $\simeq$ $19/11$) between the two cases, somewhat measuring the   relevance of the specific CA to the LI core.

    Next, let us comment on indirect measures, i.e. taken from the data in Tables 6-8: a remarkable point is the similar value
  $a_M^{(j1j2)}$  =  e.g. from 59 to  62; $\simeq$ $a_M^{p1p2)} $= e.g. from 44 to 47, for the BSS with HES as LI.

  Next, discuss  the LI and BSS, from  the (jp) point of view, see Table 8.
       \begin{itemize} 
       \item    $m_a$  has an expected hierarchy,
     \item   likely due to  the $r_M$   hierarchy factor %of 3: 592 = + = 196 or 216; 319 = + = 147 (or 40)
      \item  which is reproduced in the $A_a$   hierarchy .%LI vs. BSSs:    factor of 3:  1625 = + = 524, 469; 810 = + = 280 (or 50)
  \end{itemize}
 
 Finally  "compare" LI and BSS from the core/whole set point of view, i.e. $a_a$/$a_M$, i.e. Table 8.  First, observe that $A_a$ /$\sum$  has a quasi universal  value $\sim$  0.5. Unexpectedly, but remarkably, the ratio is close to 1/2, - with  a  j-HES minor exception, where the ratio $\sim 1/3$.  
  
{\it In fine}, the  values can be interpreted as being due to  the size of the number of CAs, i.e., $r_M$  or the cumulated  number of joint publications, i.e. $\sum$.  

 It is still emphasized that  very similar kings and queens occur in different publication types. The above findings suggest that $m_a$ is "size" dependent, - fortunately. 
 
  It is also remarkable that the $a_a^{(jp)}$ values of the LI are almost equal to the  sum of the $a_a^{(jp)}$ of the two BSS, so considered. This indicates that, at least in the case of HES and MA, the main two BSSs are the fully relevant items, -thus giving some weight to the approximation made in Sect. 3,  according to Table 3, i.e. only considering {\it only two} BSSs for, in the present paper,  such each LI.

  \section{Conclusions} \label{sec:conclusions}

The present paper aims at  generalizing the coauthor core definition (Ausloos 2013) to  collaborative work involving  some group leader and some second in charge  defined as a binary  scientific star system. The paper should end with practical considerations based on sound findings. Let us thus successively  recall ranking considerations of scientists,  the present measures on teams, and conclude with suggestions.

   Indeed, scientific quality is the most difficult and contentious item to measure. Often,  scientific production and  citations  are used   for quantifying  %as Beck \cite{beck84b}  discussed, it seems that only
      scientific achievements.  %equal in "epistemological rank" might be admitted for statistical counts 
       This has led to define a journal impact factor {\it in illo tempore}, and  has recently culminated in the $h$-index for ranking individuals,   journals, or teams and research centers.   However, one has also questioned  whether the $h-$index  is the ideal way to measure research performance (Persson et al. 2004, Bornmann \& Daniel 2009). 

       In fact, much debate occurs on how to modify the ranking due to   criteria  aking into account joint publications. It is accepted that the counting, whence ranking, is  sensitive to data size, while  interactions and contextual variations are somewhat hidden\footnote{  Let  the  very interesting work on   the critical mass and the dependency of research quality on group size by    Kenna  and  Berche   (2011) be mentioned here.}. However,  it is unlikely that teamwork might be chosen for some other  major reason than its effect on output. 

Ausloos coauthor core definition and measure tackles such considerations in a constructive way, through the relationship between the number ($J$) of (joint) publications with coauthors ranked according to their rank ($r$) importance. The approach presents a great difference  with respect to the Hirsch index  (Hirsch 2005,  2010).  The latter tests the popularity of a paper. The former emphasizes the role of persons. This seems appropriate for evaluation, and likely for weighting CAs role in quoted publications, because it
  is of common knowledge that a LI often delegates some responsibility to senior researchers in order to pursue some activity in specific fields,  often remaining in charge of each publication, while junior workers are needed for maintaining some high publication rate.  
  It has been  here above  examined what  this  BSS-like system implies on Eq. (\ref{eqZipf}) and  Eq. (\ref{eq1}).
 
 Thus a specific  test of the findings  in (Ausloos 2013), i.e.,    $ J \propto 1/r^{\alpha}$, has been made and discussed   here above  considering two prolific authors  working in the field of statistical physics.  It is confirmed that
 $\alpha\simeq 1$,  also holds for BSS systems, again when $J$  and NDCA are  "large", - such that the "statistics makes sense".. Interestingly, the  NJP, NPmfCA, NDCA, NCA, $\alpha$, $R^2$, and $m_a$ values do not seem to depend on the time regime, thus on the emphasis on one or another topics, nor on the CA.     Moreover,   it seems that $A$ and $\alpha$ do  not change much  for a given author or  even a given BSS as a function of time. Surprise, surprise.

   The role of the main coworkers, in fact  the main coauthors, has been examined, at first,  supposing that publications in peer review journals  and in so called proceedings  might have some different  influence on the core of coauthors of a LI.   Indeed, on one hand,   the two types of publications do not often rely on the same principles nor have the same goals or timing. No drastic difference has  been seen. In fact, the number of CAs is approximately the same, be they frequent CAs or accidental CAs, in particular those occurring only once, i.e. NJP1CA. This goes against common expectation.

   Observe that there is a bonus in measuring the core of CAs rather than the $h$- index of a LI.  Duplicate papers, sometimes with only cosmetic changes,  are counted several times, - appearing in peer review journals and in several proceedings-like media.  The true impact of the findings by such a LI should result from the (linear) addition of the quotations,  but that does not seem to be done, because it requests some profound reading of the publications. {\it In fine}, it lowers the $h$-index. In contrast, the $m_a$ index is  barely sensitive to the duplication, since the CAs are quasi the same on both types of publications on the same (or so)  subject, - usually with some slight difference in the CA order.    
   
   A measure $m_a$    of the relevant core of coauthors thus presented in (Ausloos 2013) has thus been of interest for checking the core measure for a BSS.
   Practical considerations on LI and research team implications can  be  first observed through the NJP exponent and coauthor core value, but introduced indirect measures are also possible  evaluation criteria. Some systematics has been found. Eq. (2) seems well obeyed. Note that if  universal values are obviously of  theoretical interest,  existing deviations  are of practical interest.
   
   A final point: whether a BSS system is useful for the net output of the members of the system is an open question, following the findings of Hollis  (2001). 
 Paraphrasing Hollis, it can be stated: {\it If society only cares about research output because of its value as a signal that the authors are competent ... , then co-authored articles may not need to be discounted.}  The metaphysical value of some cooperation, reciprocal altruism or manipulation, hypocrisy or prestige   is  reminiscent of questions on natural evolution, among which is the case of any academic career     (Petersen et al. 2011, Zhou et al. 2012). How fundamental are these questions in the present framework ? ...  One should  conclude that the  main BSSs are very relevant for a LI.  It seems worthwhile to have outstanding teams, - even with LI belonging to different research centers (like HES and SH). 
  
One may  also conjecture that
 one can introduce selection and rewarding policies in the  funding of a team (or LI)  through    $m_a$   (Ausloos 2013) and the above core and  out-of-core  measures.  It is surely unexpected that a LI needs only a couple of CAs in order to form his/her BSS basis for publishing and attaining his/her CA core.  In this respect, the effect of CA collaboration in order to increase the LI $m_a$ core value, as discussed in the Appendix,   reveals that  a cost-like matrix functional can lead to algorithmic means. 
 
 Much can still be done after the above. Three suggestions:   (i) checking the deviations from a regular hyperbolic law for BSS systems, %The effect of NDCA entices a long tail, but  a relevant observation in the present work is the king and queen effects. 
   (ii) reassessing the influence of  the NJP1CA,  i.e. somewhat outliers,  	and (iii) investigating different scientific fields.  Should one also consider more than two CAs, and look for ternary stars, and larger bodies within structures?
   %The number of coauthors seem to grow also.

  \bigskip

 {\bf Acknowledgements} \bigskip
 
 Thanks to  J. Miskiewicz  and H. Bougrine for private communications on  (Miskiewicz 2013) and   (Bougrine 2014), respectively,    prior to manuscript submission. Thanks to A. Scharnhorst and S. Wyatt for comments. This paper is part of scientific activities in COST Action TD1210, though the paper mainly stems from the defunct COST Action MP0801.
 
 \vskip 0.5cm
 
% \newpage

{\bf Appendix:  Cost model of cumulative CA core value  for a LI}% $m_a$
%\\

\bigskip
According to Fern\' andez-Cano et al. (2004),   {\it Statistical methods based on mere arithmetic counts are only partially adequate  because any quantitative bias omits relevant qualitative features and  the counting is  sensitive to data size.  Interactions and contextual variations are somewhat hidden}. Some emphasis  on group size increasing research quality  by    Kenna  and  Berche  (2006) should be also mentioned  here, since the study elaborates on a so called critical mass, in a similar line of thought as the present CA core measure in the main text. It was found that  conclude that the best group size for experimental physicists is around 25 researchers, while in theoretical physics the number is 13. Adding more researchers to the group over these sizes does not result in an increase in research quality.

  One argument about the size dependence of $m_a$, and other measures, may stem in the expectation that   %  \cite{PriceModelSgrowth2004}.
 papers, sometimes with only cosmetic changes,   reproduce similar results, but are counted in various ways, either cumulating references or a few being disregarded, because of the need for a limited bibliography.  This is a hindrance for the $h$- index, since similar papers are not cumulated for measuring the "quality" or "impact" of some work.  Since  it is  accepted  to prefer a quantitative approach, even if  approximate, {\it to any purely qualitative analysis, it is necessary to seek any data that can be obtained by a process of ''head-counting''} (Price 1956). A study of the number of coauthors  and its corresponding number of joint publications  goes in line with the usual knowledge that scientists who collaborate may bring additional  goals to a collaboration (Sonnenwald 2003).  %As she points out : a typical example is a junior scientist who wishes to be promoted and receive tenure, in addition to contributing to a collaboration. 
 
 Thus individual goals can influence a scientist  ongoing commitment to a collaboration and his or her perspective on many aspects of the work  (Sonnenwald 2007). In so doing it brings much  influence on the coauthorship list   (Kwok 2005). However, the list of coauthors for a set of results "belonging to the same research aspects, does not change much, as seen in the main text. Thus such an influence can be thought to have some aspect of prisoner dilemma game  (Bolle  \& Ockenfels 1990,  Bolton  \&  Ockenfels 2000).
 One may think that the resulting $m_a$ of all published works by a LI with his/her CAs result from a nonlinear combination of different collaborations. The result of having several CAs and/or mainly a few members of the core surely implies an increase in the $m_a$ value over the  respective $m_a$ of the CAs. it is proposed that the resulting $m_a$  occurs through a cost function based on the "adjacency matrix" made of the individual cores. This "game model" is better understood through the following examples derived from the above results on the two examined LI and their main CAs.
 
 From Table 4, it is known that the $m_a$ of HES\&SH and HES\&SB are respectively 16 and 15. In the case of MA\&RC and MA\&NV, the $m_a$ values are 11 and 5 respectively.  These can be taken as the values of the diagonal   elements of a "cost matrix"  $  \mathcal{M}$

\begin{equation} 
 \label{M}
 \mathcal{M}  =\left( \begin{tabular}{ll}   $ M_{11}$   &  $ M_{12}$   \\   $ M_{21}$   & $ M_{22}$  \end{tabular}\right)
\end{equation}

The off-diagonal elements are found as described in the text for the LI and the main CA independently of each other, except that the NJP intersection, between the two main coauthors of HES and MA, must be additionally searched for also. It  can be found that 
 $ M_{12}$  = $ M_{21}$ = 11 and  3 respectively. The eigenvalues and corresponding eigenvectors of  both  $  \mathcal{M}$ are  easily obtained. They are (26.51 and 4.49) and (12.24 and 3.76) respectively.
   %  (26.51 /12.24 = 2.17) 
   
 Recall  (see Table 2) that the overall HES-$m_a$ is equal to  26, while the  MA-$m_a$ is equal to 19.  It appears that the HES-$m_a$  value is rather more quickly reached  than the  MA-$m_a$  value  by combining, as suggested,  $m_a$ values from the best two main CAs. This can be attributed numerically to the facts that the ratios HES-$\sum$/MA-$\sum$, i.e.  3889/1554, $\sim$ 2.50   (see Table 2), and/or  the ratio HES-NJP/MA-NJP,  i.e.  1092/523, $\sim$ 2.09   (see Table 1) are large, implying a different weight of the two main CAs, thus of the two main BSSs,  in the resulting $m_a$ of the LI.  This is an  {\it a posteriori}  astounding  proof of the BSS role and of its effects.

% \newpage

%\PACS{82.\\20.Wt, 87.23.Ge,89.75.Hc   ,89.75.Da}
%{Computational modeling; simulation} \and
%{Dynamics of social systems} \and
      %}%{Networks and genealogical trees} \and
   %{Systems obeying scaling laws}   
\maketitle

    \begin{figure}
\centering
 \label{fig:Plot26hesshhessbmarcmanv} 
\caption   {Low ranking regime of the Number of Joint Publications (NJP) of  two  BSS made of a PI (HES or MA) with one of  their two main CAs (SH or SB, on one hand, RC or NV, on the other hand) as a function of the rank of the other CAs; best fits by a power law are shown together with the line indicating the threshold on how to measure  the $m_a$ CA core value
%Plot26hesshhessbmarcmanv.pdf
} \includegraphics [height=20.0cm,width=15cm]
{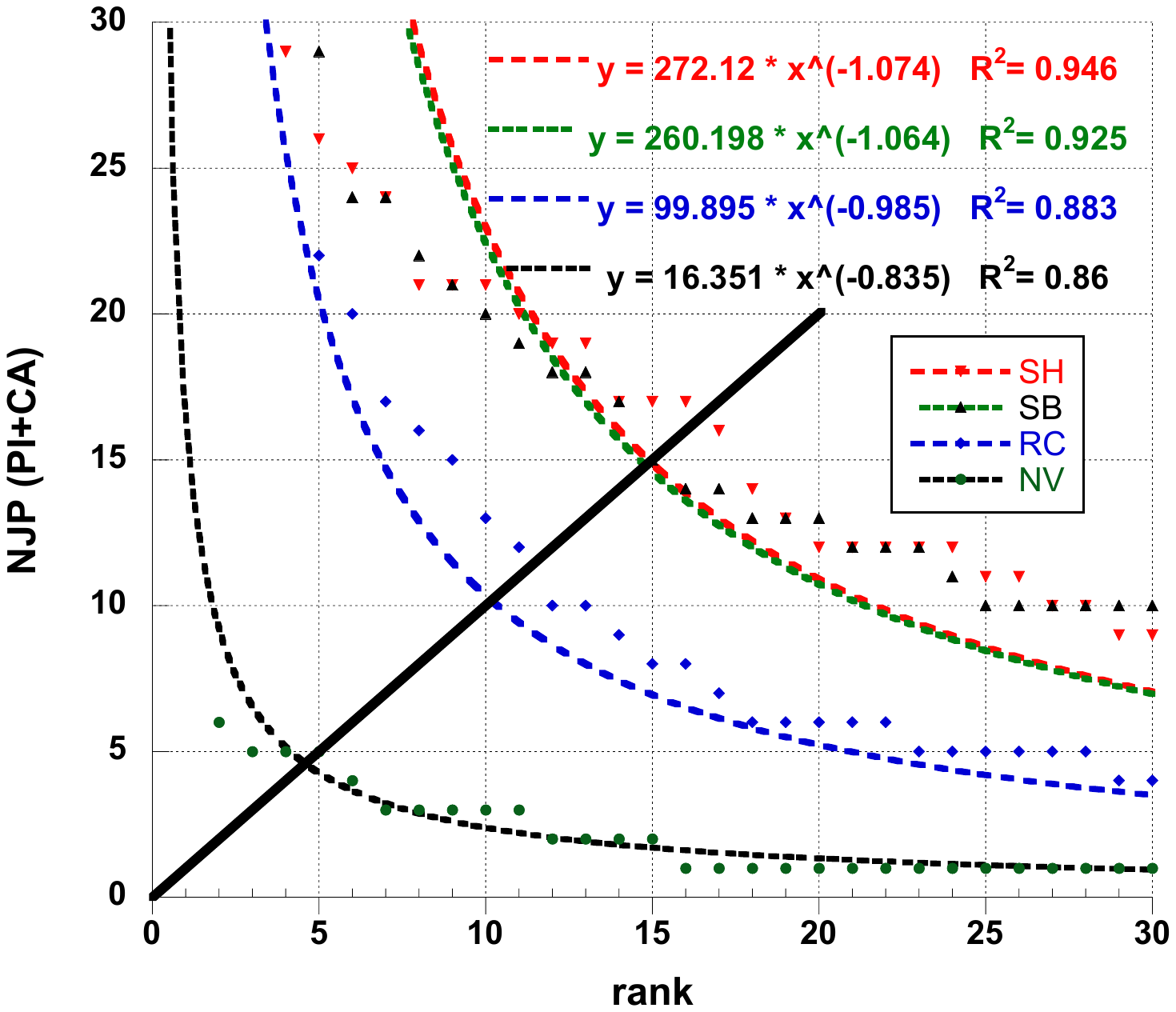}
\end{figure}

 \begin{figure}
\centering
\caption   { Number of Joint Publications (NJP) by HES\&SB as a function of the   rank of CAs  by decreasing importance.   Best fits by a power law are shown together with the line indicating the threshold on how to measure  the $m_a$ CA core value.  This allows to compare the value of Eq.(\ref{eq1})  in  time interval 1 and 2 through  p1j1 and p2j2, as well as to compare   the whole set of publications in proceedings (p1p2) with the whole set of papers in peer review journals (j1j2) } 
 \label{fig:Figabexp} 
\includegraphics[height=20cm,width=15cm]{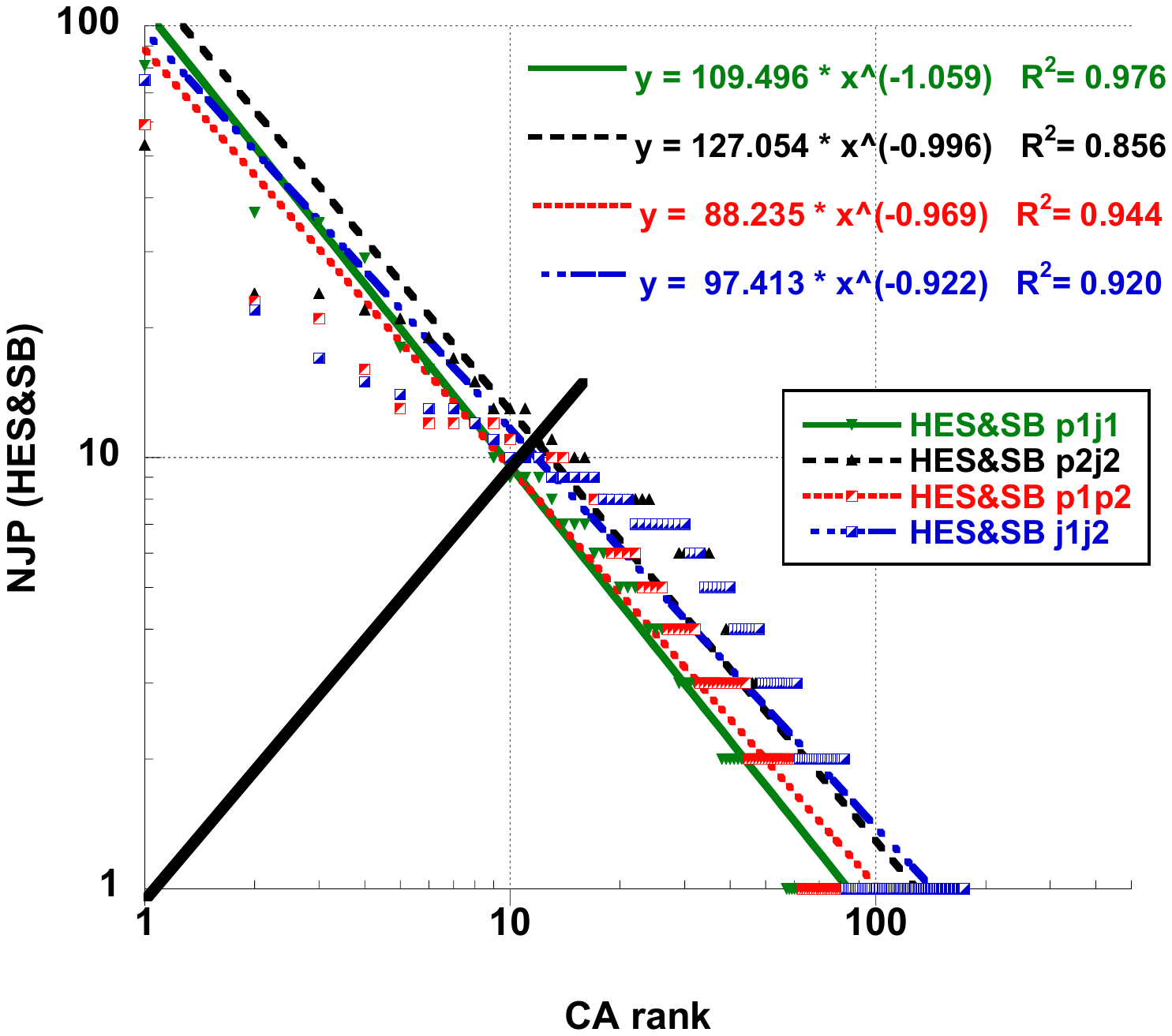}
\end{figure}

 \begin{figure}
\centering
\caption   {Number of Joint Publications (NJP) by HES\&SH  and HES\&SB  as a function of the  rank of CAs  by decreasing importance.   Best fits by a power law are shown together with the line indicating the threshold on how to measure  the $m_a$ CA core value.  This allows to compare the parameters in Eq.(\ref{eq1})  for the two main CAs of HES,  for     the whole set of publications in proceedings (p1p2) or in peer review journals (j1j2) } 
\includegraphics[height=20cm,width=15cm]{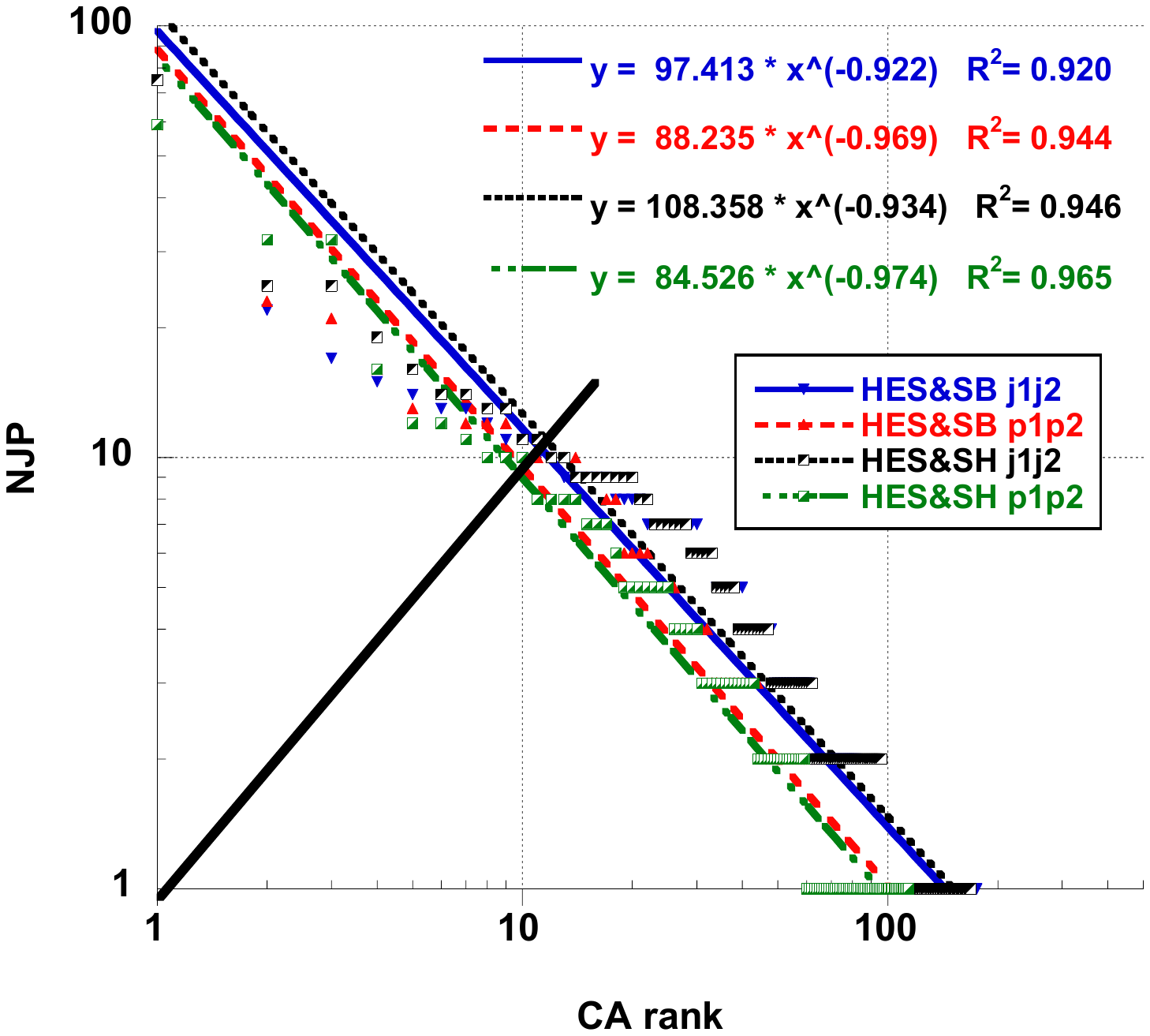}
 \label{fig:Fig1bexpbottm} 
\end{figure}

 \begin{figure}
\centering
\caption   {Number of Joint Publications (NJP) by HES\&SH  and HES\&SB  as a function of the rank of CAs  by decreasing importance.   Best fits by a power law are shown together with the line indicating the threshold on how to measure  the $m_a$ CA core value.  This allows to compare the parameters in Eq.(\ref{eq1})  for the two main CAs of HES,   in  time intervals 1 and 2 through  p1j1 and p2j2 } 
\includegraphics[height=20cm,width=15cm]{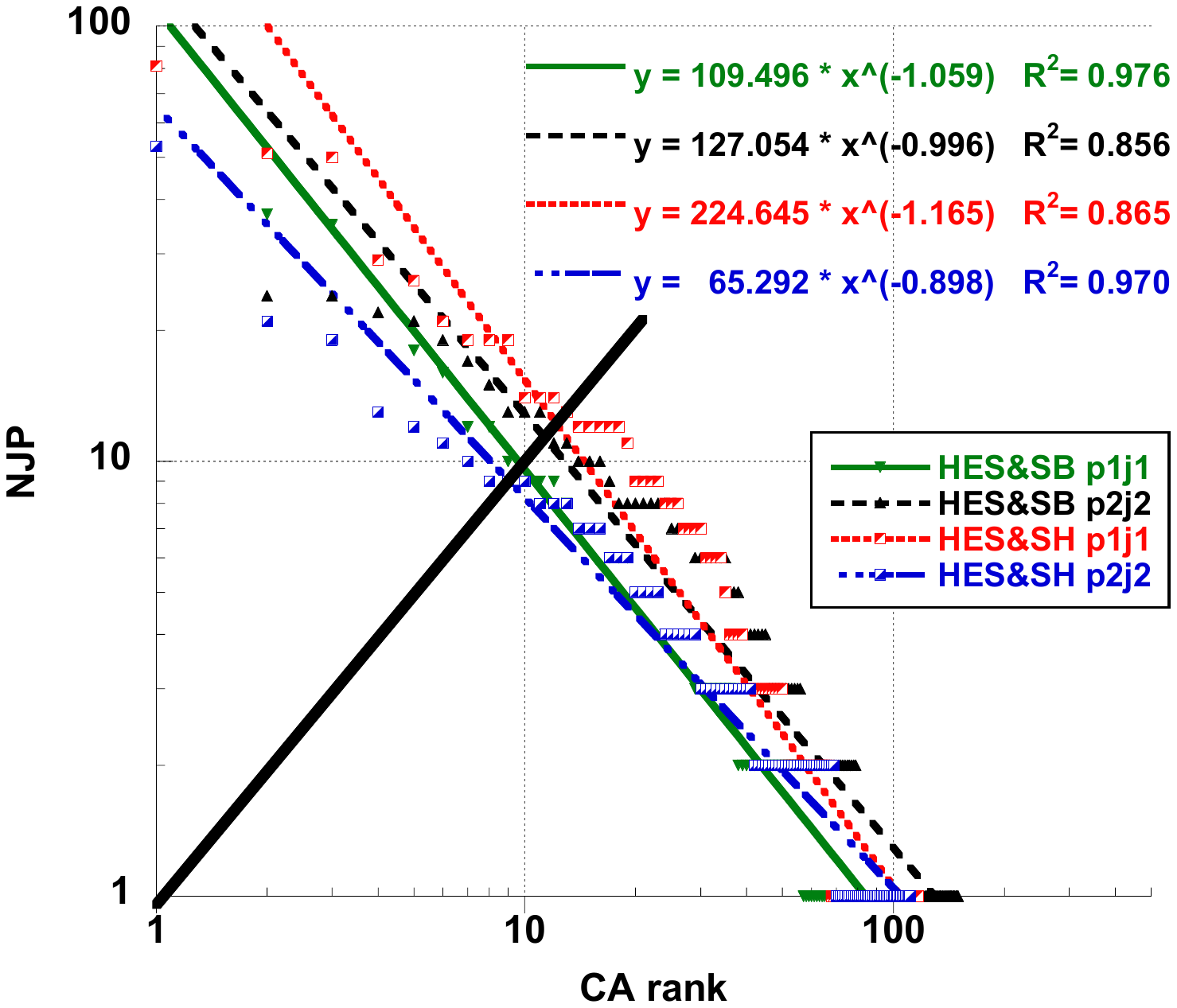}
 \label{fig:Fig1bexptp} 
\end{figure}

 \begin{figure}
\centering
\caption   {Number of Joint Publications (NJP) by MA\&NV as a function of the  rank of CAs  by decreasing importance.   Best fits by a power law are shown together with the line indicating the threshold on how to measure  the $m_a$ CA core value.  This allows to compare the value of parameters in Eq.(\ref{eq1})  in  two  time intervals, 1 and 2, through  p1j1 and p2j2, as well as to compare   the whole set of publications in proceedings (p1p2) with the whole set of papers in peer review journals (j1j2) } 
 \label{fig:Fig2aexp} 
\includegraphics[height=20cm,width=15cm]{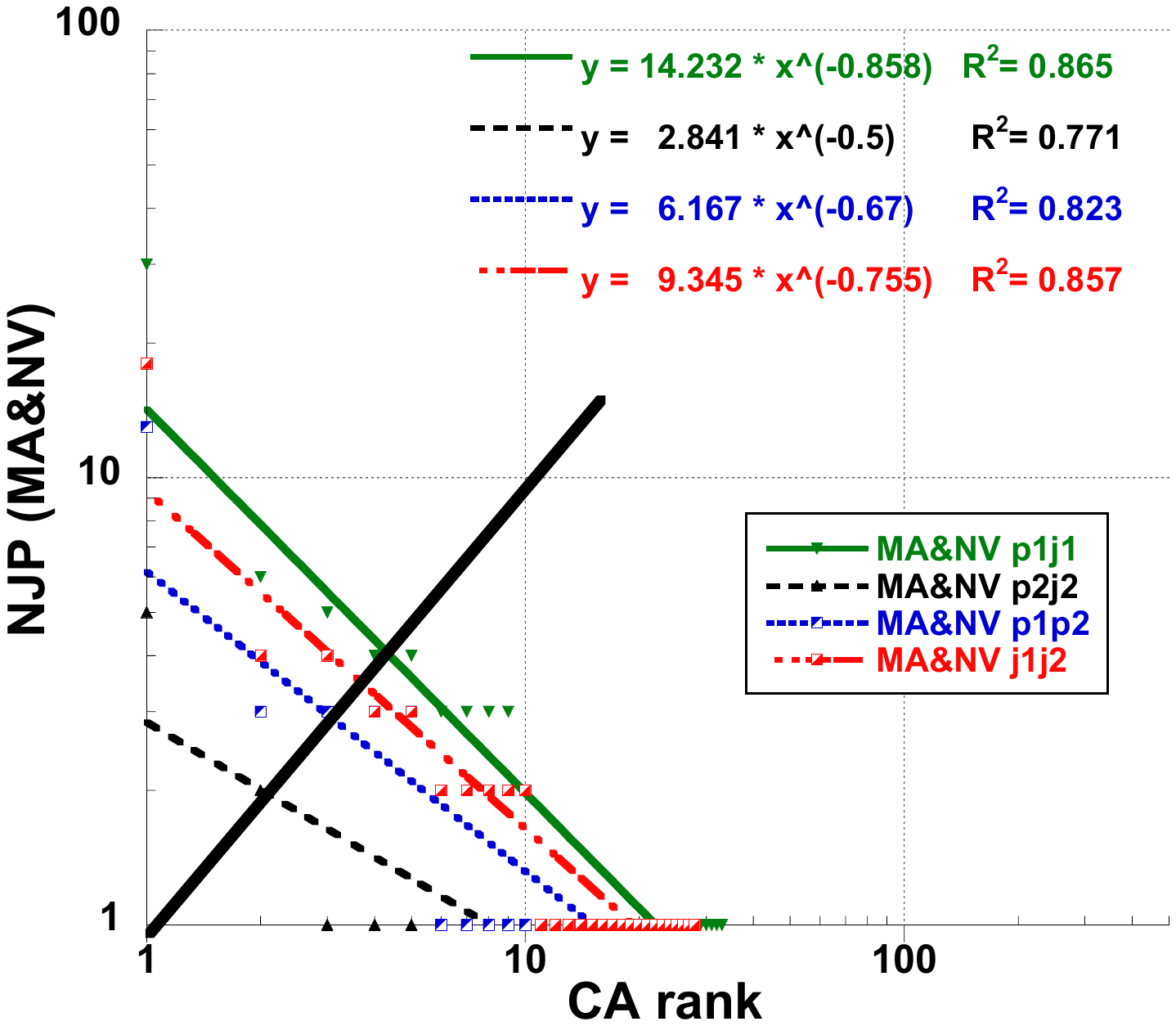}
\end{figure}

 \begin{figure}
\centering
\caption   {Number of Joint Publications (NJP) by MA\&RC  and MA\&NV as a function of the rank of CAs  by decreasing importance.   Best fits by a power law are shown together with the line indicating the threshold on how to measure  the $m_a$ CA core value.  This allows to compare the parameters in Eq.(\ref{eq1})  for the two main CAs of MA,  for     two time intervals with publications either  in proceedings (p1p2) or in peer review journals (j1j2) } 
 \label{fig:Fig2bexptp} 
\includegraphics[height=20cm,width=15cm]{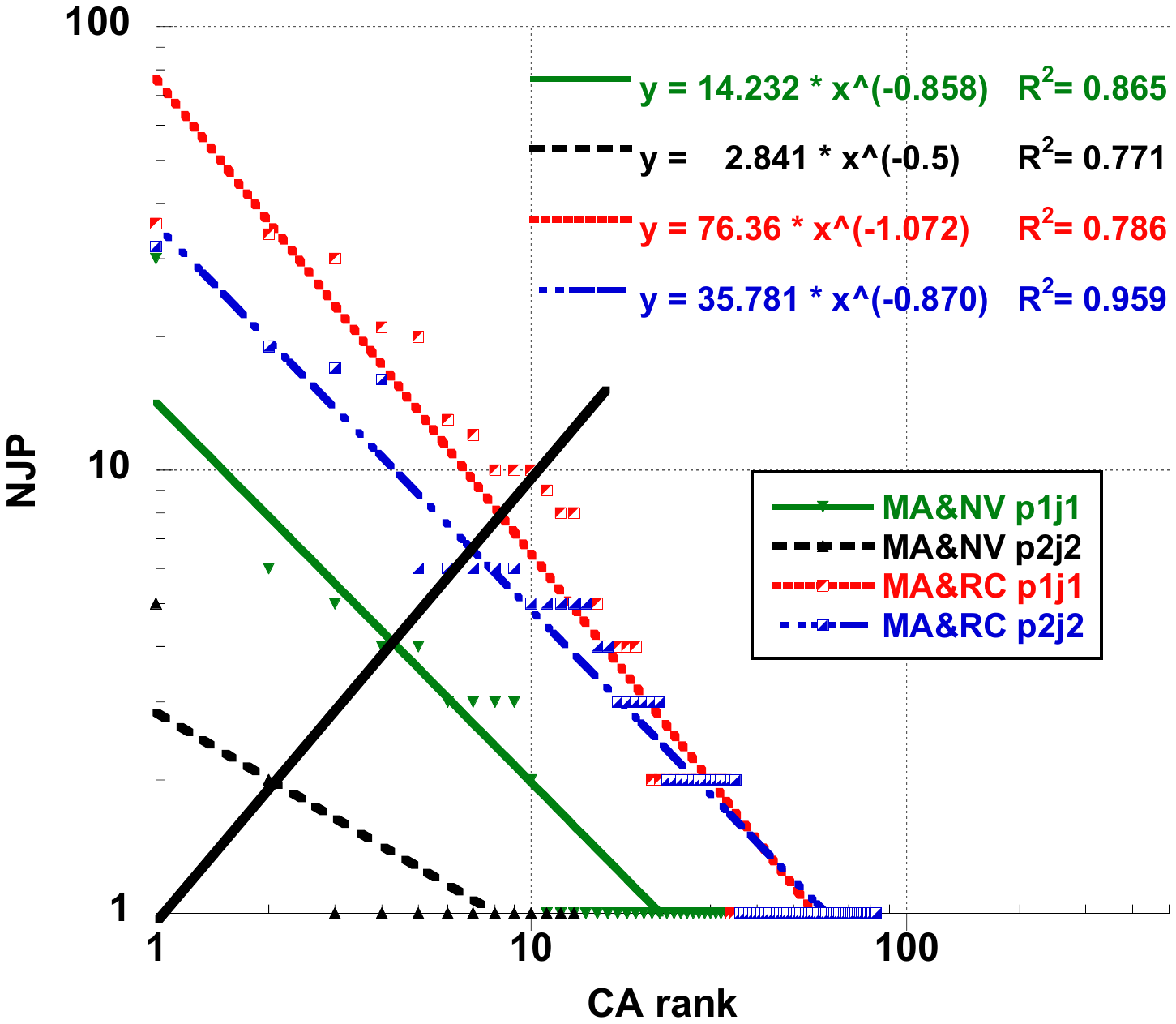}
\end{figure}

 \begin{figure}
\centering
\caption   {Number of Joint Publications (NJP) by MA\&RC  and MA\&NV as a function of the  rank of CAs  by decreasing importance.   Best fits by a power law are shown together with the line indicating the threshold on how to measure  the $m_a$ CA core value.  This allows to compare the parameters in Eq.(\ref{eq1})  for the two main CAs of MA,  for     the whole set of publications either  in proceedings (p1p2) or in peer review journals (j1j2) }   
 \label{fig:Fig2bexpbttm} 
\includegraphics[height=20cm,width=15cm]{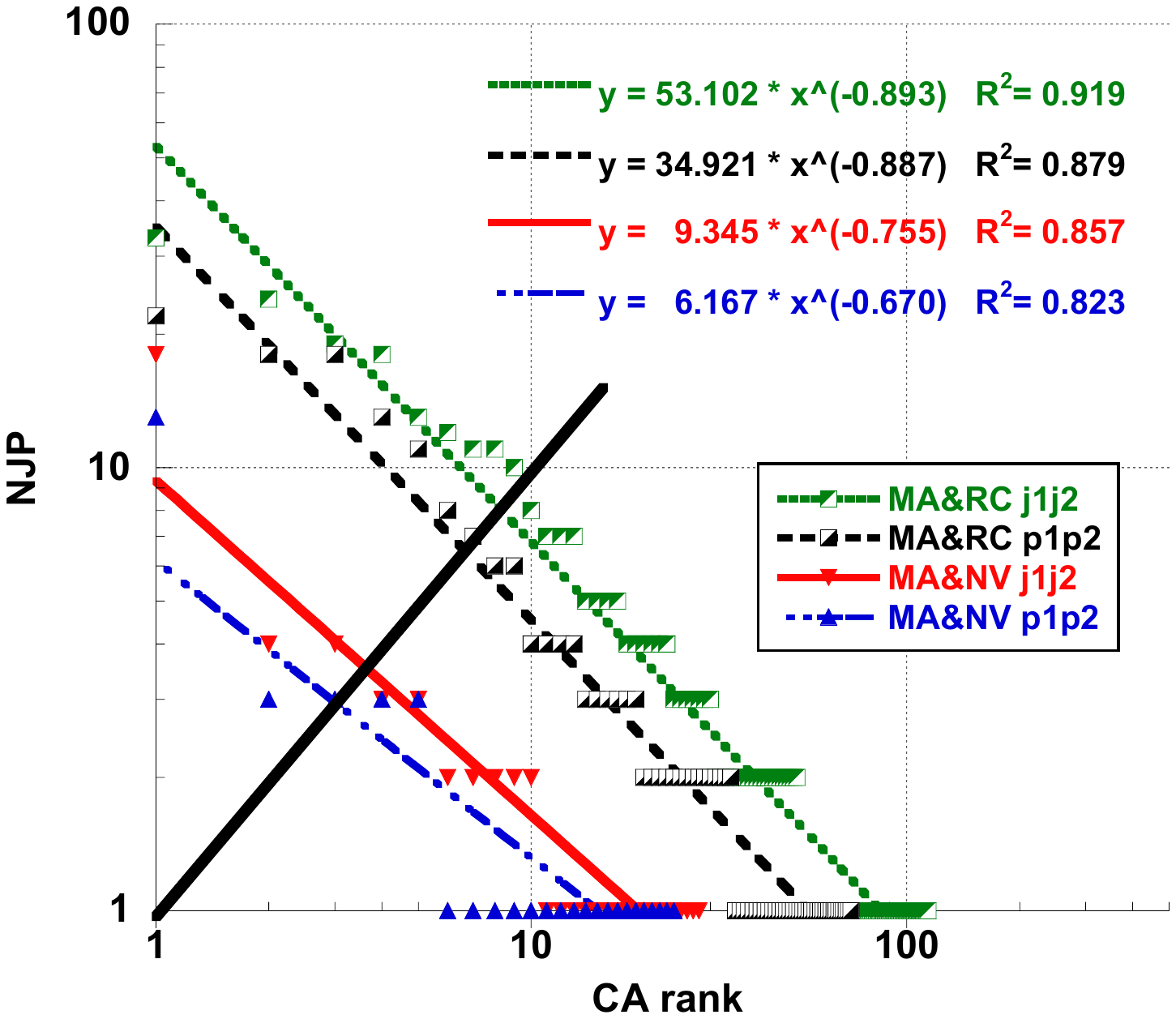}
\end{figure}

   \begin{figure}
\centering
\caption   {Search for a theoretical relationship, based on a Price-like model    Fern\' andez-Cano et al. (2004), between statistical characteristics of BSS CA distributions, i.e. $\mu$ and $\sum$ as a function of $r_M$ ($\equiv$ NDCA); the power law and the exponential (Price-like model)  fits are shown} 
 \label{fig:Plot37musum4fitslilipdf} 
\includegraphics [height=20cm,width=15cm]
{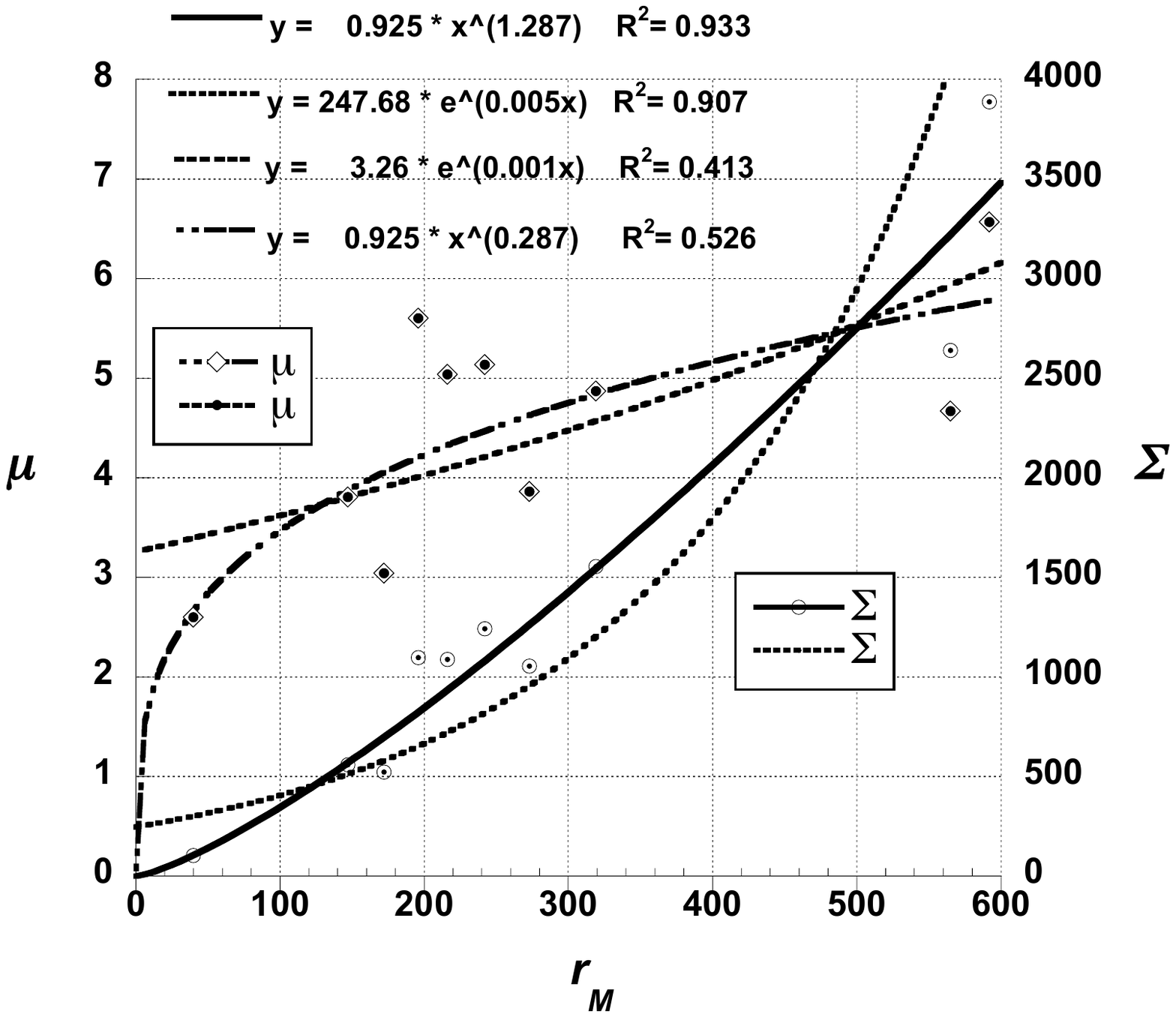}
\end{figure}

\end{document}